\documentclass[journal]{IEEEtran}

\usepackage[hyphens]{url}

\begin{document}

\title{Trust and Transparency in Contact Tracing Applications}

\author{Stacy Hobson, Michael~Hind, Aleksandra~Mojsilovi{\'c} and Kush~R.~Varshney
\thanks{S. Hobson, M. Hind, A. Mojsilovi{\'c} and K. R. Varshney are with IBM Research -- Thomas J.\ Watson Research Center, Yorktown Heights,
NY, 10598 USA, e-mail: stacypre@us.ibm.com, hindm@us.ibm.com, aleksand@us.ibm.com, krvarshn@us.ibm.com.}%
}

\markboth{}%
{Hobson \MakeLowercase{\textit{et al.}}: Trust and Transparency in Contact Tracing}

\maketitle

\begin{abstract}
The global outbreak of COVID-19 has led to focus on efforts to manage and mitigate the continued spread of the disease. One of these efforts include
the use of contact tracing to identify people who are at-risk of developing the disease through exposure to an infected person. Historically, contact tracing has been primarily manual but given the exponential spread of the virus that causes COVID-19, there has been significant interest in the development and use of digital contact tracing solutions to supplement the work of human contact tracers. The collection and use of sensitive personal details by these applications has led to a number of concerns by the stakeholder groups with a vested interest in these solutions. We explore digital contact tracing solutions in detail and propose the use of a transparent reporting mechanism, {FactSheets}, to provide transparency of and support trust in these applications.  We also provide an example {FactSheet} template with questions that are specific to the contact tracing application domain.
\end{abstract}


\IEEEpeerreviewmaketitle

\section{Motivation}
%
%

\IEEEPARstart{T}{he} recent spread of Severe Acute Respiratory Syndrome Coronovirus 2 (SARS-CoV-2) and the outbreak of the associated COVID-19 disease has inspired the development of new software applications and AI models to address many of the challenges our global society is facing. Public health agencies, corporations, and individuals have been racing to identify tools to help control the spread of the virus, find suitable treatment options, and aid in the creation of a vaccine.  Given the public health impact and urgent need to limit the continued spread of the disease, many government officials and policy makers have relaxed regulations to expedite the launch of technologies addressing these and other related concerns. Many of these technologies collect and use sensitive data about individuals such as health history, medical conditions, infection state, current health symptoms, and location. An example includes Contact Tracing Applications - those focused on identifying individuals who are at risk for developing COVID-19 through exposure to a person later identified as having been infected with SARS-CoV-2.  Contact tracing applications use various techniques to identify exposure or contact events, and use sensitive personal data like some of the examples previously identified.

The use of sensitive personal information has prompted concerns about the overall trustworthiness of these types of applications. These concerns have motivated interest in application transparency, so that application stakeholders can better understand details including the purpose of the application, the data that is collected and the application's use of the collected data.  

In recent years there has been significant discussion around the need for transparent reporting, specifically with regards to AI models and services. We apply one of the recent transparent reporting techniques in the context of contact tracing applications. Although this category of applications are not considered AI, there is significant risk to the end-users of the applications given the health implications and use of sensitive personal data. Studies have shown that technologies that are applied in a healthcare or a public health setting can lead to negative outcomes like medical errors, harm, or death especially if they are poorly designed, implemented, or applied \cite{healthIT}. The limited understanding of details of these applications motivates a need for transparency to support trust.

The objective of this paper is to identify how we develop and use transparent reporting mechanisms for Contact Tracing applications.  We do not aim to make direct conclusions about the trustworthiness of specific applications but focus on the types of questions that must be addressed to provide transparency of and support trust in the applications in this domain. 

\section{Transparent Reporting Mechanisms}
Researchers in the software engineering community have 
focused on creating useful documentation for applications. They have identified quality issues in existing documentation for conventional
systems~\cite{garousi2013evaluating,robillard2011field,sohan2017study} and discussed problems such as missing rationales for design decisions, too few examples to understand how to use a module or package, lack of overviews to illustrate how a system’s component parts work as a whole, and insufficient guidance on how to map usage scenarios to elements of an API.

AI applications pose a unique challenge, given their reliance on training data, and their often probabilistic behavior with respect to test data.  Thus, there has been a recent focus on transparent reporting mechanisms for AI systems, focusing on datasets~\cite{gebru-2018,HollandHNJC2018,data-statements}, models~\cite{Arnold2019, model-cards} and services~\cite{Arnold2019}. There have been efforts focused on the ethical development of AI that also highlighted the need for transparency or detailed assessments of AI systems \cite{EuropeanCommission2019}.

We build upon these efforts of transparent reporting to examine and provide transparency of contact tracing applications.

\section{COVID-19 and Contact Tracing}
SARS-CoV-2 poses a significant health challenge for global communities in that there are currently no identified vaccines or accepted proactive treatment methods for COVID-19, the disease the virus causes. Limiting the spread of the virus has emerged as one of the primary targets to reduce the occurrence of COVID-19, and the impact on individuals and the overburdened healthcare system in many countries.  Two of the measures used to reduce the spread are 1) limiting the physical interactions and contact between people (social distancing) and 2) identification of people who have come into contact with or proximity of an infected person (contact tracing).

Contact tracing has been used for many years as a method to control disease and has primarily relied on mobilization of trained human contact tracers - people who actively work with individuals with confirmed infections to generate a list of people whom they may have further exposed or infected \cite{CDC}.  The contact tracers then notify each of the identified individuals of the exposure risk, encourage them to get tested for infection, and suggest potential immediate quarantine action.  If any of those individuals are infected, the tracers begin the process of creating an exposure contact list for each of those people for further notification and action. 

Manual contact tracing efforts are likely not sufficient in cases where the spread of the disease has been exponential, as we have seen with SARS-CoV-2.  The initial doubling of cases in China was reported at every 6.4 days before advanced mitigation methods were employed \cite{Wu2020}. A recent publication by Johns Hopkins University Center for Health Security reports that the United States will need to add approximately 100,000 human contact tracers as part of the multi-pronged effort to manage the COVID-19 epidemic \cite{Watson2020}.

One way to scale contact tracing efforts and complement the work of human contact tracers is through the use of digital contact tracing solutions.  The United States Centers for Disease Control and Prevention (CDC) identifies two types of digital contact tracing solutions - one focused on streamlining the capture and management of data on cases and contacts, the other on using Bluetooth or GPS to track an individual’s exposure to an infected person \cite{CDC2}. The approach we use for transparency can be applied to both solution types, however, we focus our remaining discussion on the most prevalent of the application types - those that fall into the latter category. 

\subsection{Digital Contact Tracing Techniques}
There is not an agreed upon single way to achieve contact tracing; at the time of writing this paper, we identified 30 contact tracing applications available worldwide.  Many of these applications establish contact events by keeping a record of all the devices (e.g. smartphones) that come within a certain distance of one another or are in the same geographical location at the same time. Once a person has been identified as infected with COVID-19 and has indicated it in the application, a notification can be sent to all other devices running the same application that indicated close proximity to the device of the infected person within a set date range. These details are used to infer a \textit{contact event} - that one or more people were close enough to an infected individual where respiratory droplets could pass from the infected person to the others. The most common approaches for digital contact tracing rely on either the use of location tracking through the global positioning service (GPS) or Bluetooth Low Energy capabilities.  

Most smartphones today continuously capture details on the device's location and the associated time via GPS satellites.  GPS-enabled devices are reported to work best when they are outdoors under open skies where they can accurately capture location within 16 feet \cite{GPS}. Location accuracy is known to degrade when devices are indoors, underground, or near items that obstruct a direct path to the satellites, e.g. buildings, bridges, or trees \cite{GPS}.  GPS-based location tracking for a device is achieved through trilateration using radio signals from GPS satellites.  The resulting coordinates indicating geographical location are paired with a timestamp to represent location at a specific time. Contact tracing applications infer contact events by 1) identifying devices that have geographical coordinates that fall within a set distance parameter (e.g. 6 or 10 feet), 2) has a time-stamp that overlaps with one another and 3) continues to remain within the distance parameter for a specified duration (e.g. 15 minutes) even if the geographical coordinates between one or both change.  This inference also relies on certain assumptions including that the device is always in a person's possession and that possession is by a single person. A contrary case includes when the device is not in someone's possession, for example it is left somewhere (on the seat on a train, a table in a restaurant, etc.).  The device's location (and not that of the person) would be tracked and a faulty contact event can be reported.  Similarly, if a device owner or primary user lets someone else (friend, family member, etc.) use the device and the owner is later found to be infected, the exposure of others may be reported for cases where the infected individual was not present.  Additionally, the challenge posed by inaccuracies for device use indoors may limit the ability to identify a significant portion of contact events and has been identified as a potential shortcoming of using GPS location tracking for this particular purpose.

Bluetooth Low Energy (BLE) capabilities can be used to establish contact events though proximity detection. Most smartphones are equipped with bluetooth capabilities that are leveraged for this method of contact tracing and, unlike GPS, can track proximity events indoors or outdoors. Since bluetooth is used to track the \emph{promixity} to other bluetooth-enabled devices, it does not track actual location.  This has been considered one of the limitations of this method since it cannot assist in identification of geographical areas where the virus is spreading. 

In a contact tracing context, Bluetooth Low Energy is used to broadcast information from a device including a time stamp and an identifier.  Since Bluetooth Low Energy is based on short-range communications only devices that are within a short distance are expected to receive the broadcast.  The receiving device uses the received signal strength indicator to infer distance between itself and the broadcasting device. A recent tech report highlights issues with relying on signal strength as an estimator of distance.  The authors showed that the signal strength varied substantially based on the orientation of the device, absorption of the signal by the human body, and reflection or absorption of radio signals in buildings and trains \cite{Leith2020}.

Another fundamental difference between GPS-based location tracking and proximity identification through Bluetooth Low Energy is in how the data related to potential contact events are stored.  GPS-based location tracking relies on centralization of data to a remote server while the Bluetooth Low Energy technique can be either centralized or decentralized, with data being shared only locally on the individual devices.

Other, lesser-discussed methods for contact tracing solutions may involve the use of bluetooth beacons \cite{BluetoothBeacons}, location tracking through cellular or wi-fi, or tag scanning (e.g. QR or RFID). These techniques can be implemented as the sole method for contact identification or in combination with one of the other techniques.  

Although these techniques can be used to identify potential contact events, they do not factor in pertinent details that can affect transmission likelihood. For example, transmission within an indoor, poorly ventilated space may be more likely than transmission in an outdoor space \cite{Morawska2020}. Additionally, appropriate use of items like medical-grade masks or respirators by one or more individuals can greatly reduce likelihood of transmission during contact events.

We are not suggesting that one method is better than the other; we only present a brief introduction of the techniques since aspects of the technical implementation are important considerations for stakeholders interested in application transparency.   

\section{Application Stakeholders} \label{stakeholders}
Contact tracing has been identified as critical to the ability to manage the COVID-19 pandemic and, along with significant testing capabilities, may be a required item to enable governments to relax measures in place that limit the movement of their citizens \cite{Feretti2020, Kucharski2020}. These measures have negatively impacted global economies given the effect the restrictions have on businesses in industries such as retail, hospitality, and travel \& transportation.  Multiple stakeholders are interested in the development and use of contact tracing applications and the underlying motivations for this interest may differ for each group.  Understanding some of the motivations for the stakeholders provides a foundation for identifying the expected benefits and concerns of use of these applications.

\subsection{Public Health Officials}
One category of stakeholders includes those with a public health role - officials in organizations with a focus on the identification and management of viruses like SARS-CoV-2.  Examples of these organizations are the CDC in the United States and Ministries of Health in other countries.  They have an interest in digital contact tracing solutions as a complement to manual contact tracing efforts that many of them have employed for decades, with a goal of using these techniques to mitigate disease spread.

\subsection{Health Care Providers}
A second group includes health care providers - hospitals, long- and short-term care facilities, and laboratories. Their interests also include the management of the virus but extend to use of the applications for reporting of infected cases from their patients and appropriate handling of people who have been notified of potential exposure.  The health care provider and public health groups, along with government officials, likely also have interests in using applications to identify 'hot spots' or locations where the spread of the virus is growing.  This information can be used in tailoring localized measures aimed at reducing the continued spread.  

\subsection{Public and Private Companies}
Many public and private companies have interests in digital contact tracing applications as part of the efforts in allowing their employees to return to a physical worksite.  Since the start of the pandemic, many governments have instituted measures to encourage or mandate that their citizens remain at home with exceptions being allowed for those in essential roles, e.g. health care staff, public safety officials, and critical infrastructure work in specific industries \cite{CISA2020}.  Digital contact tracing applications can be used to identify exposures within an office setting and enable employers to recommend exposed individuals quarantine at home to reduce worksite-associated outbreaks.  Additionally, employees that have been notified of potential exposure through non-work related activities can communicate the exposure identification their employer and self-quarantine to prevent spread. 

\subsection{IT Professionals}
Software developers and information technology professionals are often the groups that are responsible for the development of contact tracing applications.  People within this group may have interests in developing their own solutions for contact tracing to make available to the stakeholder groups mentioned above. There are often government and geographical considerations that apply to the applications, so the potential for adoption of an existing application by a different country may require updates by software developers to make them adhere to specific local policies or regulations.

\subsection{General Public}
The final stakeholder group we identify here includes the individuals that are expected to actively use contact tracing applications.  This could be individuals in a certain country, state, or geography, or in a business context, the business' employees. Trust in the applications by the target end-users is critical for effective adoption, especially in cases where application usage is not mandated.

\section{Benefits and Concerns}
The use of digital contact tracing applications is expected to provide a variety of benefits but also brings to mind a number of concerns including those relating to privacy and security. The means in which the concerns are handled may differ given the technical design and implementation of each application. 

\subsection{Benefits}
One key benefit of contact tracing that applies to both manual efforts and digital applications is the ability to identify people who are exposed to an infected individual to encourage testing and quarantine. The implementation of a quarantine action for people who are infected but are pre-symptomatic or asymptomatic reduces the chance of them infecting others prior to awareness of their own infected state.  Recent studies have suggested the median incubation period for COVID-19 is about 5.1 days \cite{Lauer2020} and that a portion of the spread of SARS-COV-2 is from pre-symptomatic \cite{He2020, Kimball2020} or asymptomatic individuals \cite{Kimball2020, Bai2020}.  Therefore, the identification and quarantining of pre-symptomatic and asymptomatic people may be a useful factor in reducing the continued COVID-19 infections.  

Digital solutions may also provide additional benefits above that of manual contact tracing methods. Some of these additional benefits include: 

\subsubsection{Faster notification of exposure} 
Digital solutions can help reduce the time for notification to exposed individuals as compared to manual contact tracing.  Apps can notify exposed people within seconds after an infected person has been identified.  Manual tracing efforts require several steps after the identification of an infected individual, including completing an interview with the infected person or close family members to collect names of the people potentially exposed, potential additional time to locate a means to contact each person (phone numbers, addresses, etc.), and the time to establish contact.  
    
\subsubsection{Identification of contact in public spaces}
Contact tracing applications may also help address areas where manual contact tracing is not effective, for example in identifying prolonged contact with strangers or in public spaces.  A specific example of this would be an asymptomatic individual traveling via public transportation or waiting in line in a coffee shop.  The individual would not be able to identify most of the people he/she/they came into contact with.  Additionally, if the individual could recall the exact day, time, and duration of the visit, this still would not be sufficient to identify and locate all others that were in the same location at the same time.
    
\subsubsection{Identifying outbreak `hot spots'}
Contact tracing solutions that capture location details in association with infections and exposures may be useful in identifying areas where 1) infections are growing, 2) the number of cases exceed a threshold, or 3) congregations of large groups of people are enabling rapid transmission. This information may be used in implementing  countermeasures like social distancing and shelter-in-place policies targeted at specific locations to reduce the increase in infections within that geographical area.

\subsection{Concerns}
Discussions around the potential for use of digital contact tracing applications have brought light to a large number of concerns with the technologies, chief of which is focused on privacy.  We maintain that transparency of the technologies through an understanding of how each addresses the concerns is a foundation for building trust and enabling stakeholders to make decisions about which technologies they want to use and how they want to use it.  Some of the main concerns with the solutions include:

\subsubsection{Privacy}
At the core of digital contact tracing is the awareness of personal information such as health status (infected or not infected), location details, social interactions, and in some cases name, gender, age, and health history (self-reported symptoms and medical conditions).  The collection of these details pose a number of issues such as the potential for an individual's sensitive data to be made available to others (intentionally or unintentionally) and use by governments or other groups for purposes other than management of COVID-19 spread. Some practical privacy concerns are the opportunities for government agencies such as law enforcement or organizations like the United States' Immigration and Customs Enforcement (ICE) agency to surveil people through their use of the application and the potential for others to find out about their health conditions including COVID-19 infection.  
    
\subsubsection{Security}
Another top concern for application stakeholders is application security.  This includes two aspects: a) the vulnerability of the applications to attack with an attempt to change how the application works, to access personal data, or to disable usage of the application and b) the embedding of code for nefarious purposes by an application developer or publisher. The 2020 Data Breach Investigations report by Verizon identified web applications as the second highest category of healthcare industry breaches after miscellaneous errors \cite{Verizon2020}.  An example of a specific security issue with a contact tracing application was highlighted in a recent report by Amnesty International, in which they stated that they were able to access individuals' names, health status, and location details from a central server for the Qatar government-sponsored digital contact tracing application EHTERAZ \cite{amnesty2020}. 
    
\subsubsection{Coverage}
The technical implementation of the applications also affects the expectation of deployment and use.  For example, applications using Bluetooth Low Energy may require many people in the specific community or location to download and use to adequately assess potential spread amongst the population. If there is not enough coverage of use across the population, the ability to identify many of the exposed people is reduced. We understand that people may have varying reasons for choosing to participate or not, one of which is their belief of trustworthiness of the applications based on many of the specific concerns highlighted here. 
    
\subsubsection{Access}
A key requirement for digital contact tracing is that individuals have devices (e.g. smartphones) that enable the application to function properly. Since many of the applications rely on BLE or GPS, individuals would have to have devices that have the capabilities embedded. Results of a 2019 survey showed that approximately 53\% of the people aged 65 or older in the United States have a smartphone while ownership of those between ages 18 and 49 was greater than 90\% \cite{PewOwnership2020}. Also, for some of the systems, a newer version of a smartphone is required; people with older smartphones may not have the ability to use or get alerts from these types of applications. Since the identification of contact events for individuals are based on these devices, children and disadvantaged groups may also be omitted given a lack of access to or continued use of a personal smartphone. Countries like India and Indonesia have large portions of the population that either do not have access to a compatible device or have a device at all \cite{PewGeneral2020}. 
    
\subsubsection{Accuracy}
We introduced some of the issues related to accuracy earlier in the discussion, specifically the limitations with tracing contact in large locations (e.g. apartment buildings) and areas where people are more geographically separated. We highlighted a concern with GPS previously in that it is not as accurate indoors or in areas where there isn't an unimpeded path to open skies.  With BLE, accuracy may degrade based on the positioning or obstruction of the bluetooth enabled device and this may impact the proximity identification \cite{Leith2020}.  
    
\subsubsection{Asynchronous contact events}
There is potential for exposure and spread of the virus from cases where there is an asynchronous contact event, for example with a person being in a small enclosed space (e.g. elevator) for a period of time then leaves, and then shortly thereafter another person comes into the same space. There is the belief that most of the spread of the virus is through aspiration of respiratory droplets however there is also the possibility that spread occurs when an uninfected person touches an object or surface that an infected person has previously touched and then puts their hands or fingers in the areas around their mouths, nose, or eyes. Both of these examples of spread can occur through an asynchronous contact event but may not be captured as such in digital contact tracing solutions that focus on people being in the same location or close proximity at the same time.
    
\subsubsection{Device impacts} 
Each contact tracing application may also have specific considerations and impacts on the devices in which they are being run. There is a concern with the potential of high consumption of battery power with bluetooth-based techniques \cite{AppleBluetoothDocumentation2020}.  Some of the applications have requirements to run in the foreground of the device, meaning that when other applications are being used by the device holder, the application may not be able to work appropriately to identify contact events.  Additionally, the device makers may have restrictions in place that effect the way the applications work.  One example of this is Apple's restriction on allowing bluetooth transmissions when an iOS based device is locked \cite{AppleBluetoothDocumentation2020}, which limits the functionality of contact tracing applications on these devices.  
    
\subsubsection{Ability}
These applications rely not only on adoption by individuals but also appropriate use.  If people are unaware of specific requirements for use, or are not comfortable with usage of the device or the application, their interactions may not be sufficient to enable effectiveness of the application. Consider an example where a novice technology user has a smartphone and downloads the application on it.  The user may not realize that downloading the application is not sufficient, but may also require completion of a profile and providing consent for the application to run on the user's device.  Consent may also be required in the device settings to allow the application to access some of the smartphone's capabilities that are required.  For example, a user may install an application but inadvertently restrict the ability for it to work by disabling access to the device's location services or bluetooth capabilities. 
    
\subsubsection{Interoperability}
Limitations associated with contact tracing applications' ability to identify contact events may lead to missed episodes of exposure and potential transmission of the virus.  We have highlighted some of these concerns relating to the coverage, access, and accuracy aspects already. Another related concern of the application's ability to identify contact is that of interoperability between applications and/or devices.  Consider an example where an infected individual is located near another individual for an extended period of time.  If the two people are running different contact tracing applications, or running applications on different devices (e.g. one with an Android based device and the other with an iOS based device) restrictions in the applications being able to share details with one another or from one platform to another is a direct inhibitor to the identification of this contact event.  Apple and Alphabet (Google's parent company) have proposed a framework that allows interoperability between the device operating systems of contact tracing applications, which is a helpful step in addressing this issue, but is limited to the applications that use the framework \cite{AppleGoogle2020}.  In some cases, applications like Aarogya Setu have developed both a version based on the Android and the iOS operating systems \cite{AarogyaSetu2020}.
    
\subsubsection{Reluctance in disclosure}
In some cases people may agree or are mandated to use a digital contact tracing application but have an interest in withholding an infection diagnosis because of privacy or security concerns, or personal reluctance to acknowledge the diagnosis. Similarly, people may not want to acknowledge or disclose their exposure to infected individuals. In some geographies, people who are diagnosed as infected or are identified as having been exposed to an infected individual may be told to quarantine for a period of time. These measures will limit people's movements and ability to do things that they may want to do e.g. go to work, go to the grocery store, visit family members, or participate in social activities.  Some of these limitations may have an economic impact (restricting ability to work) which may reinforce a reluctance for an individual to disclose infection or exposure.

\section{Current Contact Tracing Applications}
The urgent global need for contact tracing has spurred the development of many digital solutions.  To date, we have identified 30 different applications created since December 2019 specifically to support the contact tracing needs required for management of COVID-19.  These solutions may differ in technical implementation and specific policies of use. It is likely hard for public health agencies and government officials to quickly identify the differences between applications as they try to determine which one to select as part of their targeted virus management strategy.  Similarly, it is also difficult for individuals who are asked to install and use the applications to get consumable details regarding specific considerations relevant to them like requirements for use, types of data collected, and data use policies. 

We provide a list of the 30 applications in Table~\ref{table:contacttracingapps} including details on the organization that sponsored the development or group that directly developed the application, and the technical approach that is used for identifying contact. These details are based on information reported for each of the applications at the time of authorship of this paper, but we acknowledge that due to the dynamic nature in the development of these applications and efforts to address emerging concerns of the intended community of use, some of these details may change in the future.

\begin{table}
\caption{Examples of Contact Tracing Applications}

\begin{tabular}{|| p{2 cm} | p{3.4 cm} | p{2 cm} ||}
 \hline
 \textbf{Application} & \textbf{Developer or Sponsoring Organization} & \textbf{Tracing Technique} \\ 
 \hline\hline
 Aarogya Setu & Government of India & GPS \\ 
 \hline
 Apturi Covid & \emph{Private Developers} (Latvia) & Bluetooth \\ 
 \hline
 Corona100m & \emph{Private Developer} (South Korea) & GPS \\
  \hline
 coEpi &  \emph{Private Developer} (United States) & Bluetooth \\
 \hline
 Coronika & Kreativzirkel (Germany) & Manual \\
 \hline
 COVA Punjab & Government of Punjab (India) & GPS \\
 \hline
 CovidSafe & University of Washington (United States) & Bluetooth \\
 \hline
 CovidSafe & Australian Government (Australia) & Bluetooth \\
 \hline
 Covid Watch & \emph{University consortium} (United States) & GPS \\
 \hline
 EHTERAZ & Ministry of Interior (Qatar) & GPS and Bluetooth \\
 \hline
 eRouška & Czech Ministry of Health and Hygiene (Czech Republic) & Bluetooth \\
 \hline
 HaMagen & Israel Ministry of Health (Israel) & GPS \\
 \hline
 Immuni & Italian Central Government (Italy) & Bluetooth \\
 \hline
 Ito & \emph{Private consortium} (Germany) & Bluetooth \\
 \hline
 Health Code & Alibaba (China) & GPS \\
 \hline
 Social Monitoring & Infogorod (Russia) & GPS  \\
 \hline
 Mahakavach & Government of Maharashtra (India) & GPS \\
 \hline
 NHSX & NHS Digital (UK) & Bluetooth \\
 \hline
 NOVID & Carnegie Mellon University (United States) & Bluetooth \\
 \hline
 Private Kit: Safe Paths & MIT (United States) & GPS\\ 
 \hline
 ProteGO & Ministry of Digital Affairs (Poland) & Bluetooth \\ 
 \hline
 Rakning C-19 & Iceland’s Department of Civil Protection and Emergency Management & GPS \\ 
 \hline
 Smittestopp & Norwegian Institute of Public Health (Norway) & GPS \\ 
 \hline
 StopCovid & Government of France (France) & Bluetooth \\ 
 \hline
 StopKorona! & Ministry of Health (North Macedonia) & Bluetooth \\ 
 \hline
 Stopp Corona & Austria Red Cross (Austria) & Bluetooth \\ 
 \hline
 Swiss Covid & Federal Office of Public Health (Switzerland) & Bluetooth \\
 \hline
 TraceTogether & Government Technology Agency (Singapore) & Bluetooth \\ 
 \hline
 Triax & \emph{Private consortium} (United States) & Tag Scanning \\ 
 \hline
 ZeroBase & ZeroBase Foundation (United States) & QR Code Scanning \\ 
 \hline
\end{tabular}
\newline
\label{table:contacttracingapps}
\end{table}

\section{FactSheet Template for Contact Tracing}
As a follow-on to our prior work on the use of FactSheets for transparent reporting \cite{Arnold2019}, we now aim to help identify the questions that would provide useful and critical information about contact tracing applications.  

To achieve this goal, we first compiled questions that are relevant to provide basic information relative to any model, service or application.  These include questions focused on scope of use, target stakeholders, and data that is collected.  Then, after detailed review of contact tracing technologies and their potential for use, we augmented the initial list with questions specific to contact tracing, namely, those that would elicit details addressing the benefits and concerns identified above.  Some of these questions focus on the technical implementation including the technique used for establishing proximity and/or location, method for identifying a contact event (centralization versus decentralization), and method for infection reporting.  As a final step, we considered the beneficiaries of the applications and questions that would be of interest to them that were not already identified. Examples of these questions include how infections are reported, whether usage is voluntary or mandated, and how compliance with local laws or regulations is achieved.  

These efforts enabled us to create a \emph{FactSheet template} - a list of questions that can be used to provide important details on and promote transparency of contact tracing applications. The {FactSheet} template we created is organized into four main categories: General Questions, Data-specific Questions, Privacy Questions, and Use Questions.  We introduce the template and the associated questions in Tables \ref{table:general}--\ref{table:privacy}.

\begin{table}[h]
\caption{Contact Tracing FactSheet Template: General Items}
\label{table:general}
 \begin{tabular}{| p{8cm} |} 
 \hline
 \textbf{General} \\
 \hline\hline
 \emph{What is the scope of use of the application?} \\ 
 \hline
  \emph{Who are the target stakeholders or beneficiaries of the application - the people who will be impacted by its success or failure (e.g. government or public health agencies, private companies, and/or individuals)?} \\
 \hline
 \emph{What policies or laws apply to the development, deployment or usage of this application? How do you ensure compliance with these regulations?} \\
 \hline
 \emph{Is this application intended for stand-alone use or as a companion to established health-agency or government manual tracing efforts?}  \\
 \hline
 \emph{Does this application connect to any other applications or IT systems (for example, public health, clinical laboratory, or hospital systems)?}  \\ 
 \hline
 \emph{Identify the technique used for establishing contact (bluetooth, location tracking via GPS, etc.)} \\
 \hline
 \emph{What are the specific requirements for efficacy of tracking and contact identification?} 
 \newline
    \emph{Distance – the span of space that is used to identify a contact event}
    \newline
    \emph{Time – amount of time individuals are within the required distance to meet threshold for exposure risk}
    \newline
    \emph{Coverage – the number of people or percent of population needed to use the app} \\
 \hline
 \emph{What concerns (positive and negative) might the beneficiaries have in how the service works?  How are these concerns addressed?}  \\ 
 \hline
 \end{tabular}
\end{table}

\begin{table}[h]
\caption{Contact Tracing FactSheet Template: Data Items}
\label{table:data}
\begin{tabular}{| p{8cm} |}
 \hline
  \textbf{Data} \\
  \hline\hline 
  \emph{What data is collected by the application? Include data collected directly by the app, from the user, and data accessed from other applications/system.} \\
  \hline 
  \emph{Is this data combined with any additional details about an individual, community, locale, or environment?} \\
  \hline 
  \emph{Identify any data collected that is of a sensitive nature (for example, health conditions, symptoms, etc.)} \\
  \hline 
  \emph{How is the collected data used?} \\
  \hline 
  \emph{Who has rights to access the data (explicitly define people, agencies, and/or organizations)} \\
  \hline 
  \emph{What is the policy on data retention and deletion?} \\
  \hline 
  \emph{Is there potential for the data provided or collected to be used for future purposes, beyond the scope of the current intended use?   What mechanisms do you implement to limit use beyond scope of the intended purpose?} \\
  \hline 
  \emph{What mechanisms are used to keep the data secure?} \\
  \hline
\end{tabular}
\end{table}

\begin{table}[h]
\caption{Contact Tracing FactSheet Template: Use Items}
\label{table:use}
\begin{tabular}{| p{8cm} |} 
 \hline
  \textbf{Use} \\
 \hline\hline
 \emph{What are the device requirements for use of the application? (for example, required platform, operating system, wifi, and/or cellular access, date of manufacture)} \\ 
 \hline
 \emph{Is use of this application voluntary (opt-in) or mandatory?
  \newline If mandated, do users have the ability to opt-out?
  \newline If users opt-out, what is the policy on deletion of details on their use and associated data from the system?} \\ 
 \hline
  \emph{Is user consent collected for the use of the application?} \\ 
  \hline
  \emph{Is user consent requested for access to or collection of the explicit user data (personal, health, and/or location-related details)?} \\ 
  \hline
  \emph{Are contact episodes identified in a decentralized (locally on each device) or centralized (remotely through a server) manner?} \\ 
  \hline
  \emph{How is infection being reported - self-reported or reported from an established health system (public health, clinical laboratory, hospital, or other COVID-19 management system)?  
  \newline If self-reported, how does the user indicate infection?  Is the identification by the user authenticated in some way?
  \newline If reported from an established health system, how is the information shared and received? } \\ 
  \hline
  \emph{What is the expected impact on the devices that use this app? (battery use, compute, and bandwidth considerations)} \\ 
  \hline
  \emph{What are specific considerations for use of the particular application? Include details on any technical concerns or shortcomings.} \\ 
  \hline
  \emph{What are the limitations of use? List scenarios for which use is not suitable (e.g. incompatibility with certain devices, inability to identify non-contact barriers like walls separating locations within a building)} \\ 
  \hline
  \end{tabular}
\end{table}

\begin{table}[h]
\caption{Contact Tracing FactSheet Template: Privacy Items}
\label{table:privacy}
\begin{tabular}{| p{8cm} |}
 \hline
  \textbf{Privacy} \\
  \hline\hline 
  \emph{Did you implement the right for a user to 1) withdraw consent, 2) object, and 3) be forgotten in the application?} \\ 
  \hline
  \emph{Does the application allow people to learn any personal information about others?} \\ 
  \hline
  \emph{Are privacy-preserving techniques incorporated in the application (e.g. data anonymization, encryption, aggregation)?  If so, provide details on the techniques used.} \\ 
  \hline
  \emph{What additional measures are used to protect the data and identity of infected and exposed individuals? } \\ 
  \hline
  \emph{Could this application be used in a way that identifies people who are infected or at risk to 1) the developers, 2) people within an individual's social circles, 3) to those the app is warning about contact and potential exposure, or 4) to the government, employer, or managing organization?} \\ 
  \hline
  \emph{If the app connects to public health or hospital systems, how do you ensure that personal information isn’t accessible during data sharing points?} \\ 
  \hline
\end{tabular}
\end{table}

\section{Discussion}

We have presented a broad {FactSheet} template to support transparency of contact tracing applications. A key component of {FactSheets} is the tailoring of the questions within the {FactSheet} template to address a specific stakeholder group and provide clarity on the aspects of the applications that they are most concerned about. As we discussed in Section ~\ref{stakeholders}, the motivations of interest in the applications may differ for each group, and these motivations influence the questions that enable transparency for each group.

Let's consider the General Public stakeholder group whose interest in transparency may be most related to their own use of the applications.  The questions that focus on data, privacy and device requirements may be the ones that are critical for their specific version of the FactSheet template. Some of these questions would include those relating to 1) types of data collected, 2) how the data is used, 3) requirements for efficacy of tracking and contact identification, 4) expected device impacts, 5) limitations of use, and 6) data privacy. 

The Public Health official group would potentially be engaged in the selection and management of the contact tracing applications for a specific geographical area (city, county, state, county, etc.) and therefore would likely have interest in the broadest set of questions from the template above. The full set of questions we identified in the template would provide information pertaining to the concerns from all stakeholder groups, and this could be useful as the Public Health officials evaluate and select the applications with the other stakeholders concerns in mind. However, they might be less interested in questions relating to specific device impacts or requirements of use unless these considerations could greatly limit acceptance in their geographical communities.  

For the IT Professionals group, questions around the technical implementation, limitations, connections to other IT systems, all data aspects (collection, policies, access rights, retention, and security), device requirements, decentralization versus centralization, and privacy-preserving techniques would be of particular interest. This group may also be interested in additional technical details about the applications including access to the code base.  Recent reports have suggested that the code for two of the applications referenced in Table~\ref{table:contacttracingapps} - Aarogya Setu and TraceTogether - will be publicly available as open source projects \cite{AarogyaSetuCode2020, OpenTrace2020}. We believe that this is another path for promoting transparency of these types of applications for this specific stakeholder group, and can be used together with {FactSheets} to foster trust. 

We acknowledge that there could be additional relevant questions that were not listed in our {FactSheet} template that might be useful for application transparency in this context.  We suggest the {FactSheet} template as a useful starting point in the efforts towards transparency.

We also acknowledge that as applications are updated, the answers to the questions may change.  We suggest the generation of a {FactSheet} for each application deployment and update.  It is possible to create a base {FactSheet} for the application that covers the details that will not change from one application instance to the other, and also include a supplementary FactSheet that is generated for each version or use case.

We have demonstrated the potential of {FactSheets} in this context to promote transparency but note that {FactSheets} are not limited to this purpose alone.  They can be leveraged as a mechanism in additional contexts, for example as part of a robust trust and governance strategy within a business, or a path for evaluation and certification of models or services by a third-party.  

\section{Conclusion}

Our proposal of the use of {FactSheets} for transparency will help in providing consumable details about the applications for the stakeholder groups we discussed in Section III and help each group to understand application details related to the concerns of their group.  

We encourage people with an interest in fostering trust in models, services and applications to use transparent reporting techniques like {FactSheets} to provide consumers and stakeholder groups with the necessary details to better understand these technologies.

\section{Acknowledgments}

The authors would like to thank Marc Stoecklin for input on the list of current contact tracing applications. 

\ifCLASSOPTIONcaptionsoff
  \newpage
\fi

\bibliographystyle{IEEEtran}
\bibliography{contacttracing}

\end{document}